# Linear Programming Contractor for Interval Distribution State Estimation Using RDM Arithmetic

VietCuong Ngo, Wenchuan Wu, *Senior Member*, *IEEE*

*Abstract--* State estimation (SE) of distribution networks heavily relies on pseudo measurements that introduce significant errors, since real-time measurements are insufficient. Interval SE models are regularly used, where true values of system states are supposed to be within the estimated ranges. However, conventional interval SE algorithms cannot consider the correlations of same interval variables in different terms of constraints, which results in overly conservative estimation results. In this paper, we propose a Linear Programming (LP) Contractor algorithm that uses a relative distance measure (RDM) interval operation to solve this problem. In the proposed model, measurement errors are assumed to be bounded into given sets, thus converting the state variables to RDM variables. In this case, the SE model is a non-convex model, and the solution credibility cannot be guaranteed. Therefore, each nonlinear measurement equation in the model is transformed into dual inequality linear equations using the mean value theorem. The SE model is finally reformulated as a linear programming contractor that iteratively narrows the upper and lower bounds of the system state variables. Numerical tests on IEEE three-phase distribution networks show that the proposed method outperforms the conventional interval-constrained propagation, modified Krawczyk-operator and optimization based interval SE methods.

## NOMENCLATURE

*Measurement*

| | |
|---|---|
| $P_i^{(m)}$ | Active injection power of bus $i$ of the distribution systems |
| $Q_i^{(m)}$ | Reactive injection power of bus $i$ of the distribution systems |
| $U_i^{(m)}$ | The voltage magnitude of bus $i$ of the distribution systems |
| $V_i^{(m)}$ | Square of voltage of bus $i$ of the distribution systems |
| $I_{ij}^{(m)}$ | Current magnitude of line $ij$ of the distribution systems |
| $L_{ij}^{(m)}$ | Square of current magnitude of line $ij$ of the distribution systems |
| $P_{ij}^{(m)}$ | Active power flow of line $ij$ of the distribution systems |
| $Q_{ij}^{(m)}$ | Reactive power flow of line $ij$ of the distribution systems |
| $v_*$ | The measurement error of the distribution systems |
| $\alpha_*^{(m)}$ | RDM variables corresponding to each measurement |

*Variables*

| | |
|---|---|
| $e_i$ | The real parts of the voltage at node $i$ of the distribution systems |
| $f_i$ | The imaginary parts of the voltage at node $i$ of the distribution systems |
| $I_{ij}^{re}$ | The real parts of the branch current $ij$ of the distribution systems |
| $I_{ij}^{im}$ | The imaginary parts of the branch current $ij$ of the distribution systems |
| $V_i$ | Square of voltage magnitude of bus $i$ of the distribution systems |
| $\alpha_*$ | RDM variables corresponding to each variable |

*Parameters*

| | |
|---|---|
| $s=(a,b,c)$ | Subscript indicates the phase of the distribution systems |
| $(r_{ss})_{ij}$ | the branch resistance between the phases of line $ij$ of the distribution systems |
| $(x_{ss})_{ij}$ | the branch reactance between the phases of line $ij$ of the distribution systems |
| $\overline{P_i^{(m)}}, \underline{P_i^{(m)}}$ | Upper/lower active injection power bound of bus $i$ of the distribution systems |
| $\overline{Q_i^{(m)}}, \underline{Q_i^{(m)}}$ | Upper/lower reactive injection power bound of bus $i$ of the distribution systems |
| $\overline{V_i}, \underline{V_i}$ | Upper/lower bound for the square of voltage magnitude of bus $i$ of the distribution systems |
| $\overline{L_{ij}^{(m)}}, \underline{L_{ij}^{(m)}}$ | Upper/lower bound for the square of current magnitude of line $ij$ of the distribution systems |
| $\overline{P_{ij}^{(m)}}, \underline{P_{ij}^{(m)}}$ | Upper/lower active power flow bound of line $ij$ of the distribution systems |
| $\overline{Q_{ij}^{(m)}}, \underline{Q_{ij}^{(m)}}$ | Upper/lower reactive power flow bound of line $ij$ of the distribution systems |
| $\overline{e_i}, \underline{e_i}$ | Upper/lower bound for the real parts of the voltage at node $i$ of the distribution systems |
| $\overline{f_i}, \underline{f_i}$ | Upper/lower bound for the imaginary parts of the voltage at node $i$ of the distribution systems |
| $\overline{I_{ij}^{re}}, \underline{I_{ij}^{re}}$ | Upper/lower bound for the real parts of the branch current $ij$ of the distribution systems |
| $\overline{I_{ij}^{im}}, \underline{I_{ij}^{im}}$ | Upper/lower bound for the imaginary parts of the branch current *ij* of the distribution systems |
| $\overline{v}, \underline{v}$ | Upper/lower bound for the prior of the measurement error |

I. INTRODUCTION

*A. Background*

AS few real-time measurements are available in distribution networks, implementation of the advanced analysis and optimization modules of distribution management systems (DMS) is restricted. The most widely used distribution network state estimation (DNSE) technique is the weighted least squares method [1]. DNSEs are based on the core principle that a form of normalization can be used to determine the distance between estimated and measured values; a short distance indicates that the estimate was accurate. However, the observation of most DNs cannot depend merely on real-time measurements. Therefore, DNSEs that feature pseudo-measurement generation and modeling are often used [2], which may involve significant errors. So, it is necessary to use a non-deterministic SE model to replace traditional deterministic SE models in this condition.

Interval state estimation (SE) [3] is used to estimate the uncertainty range of states in which the 'true' states are certain to be found. This guaranteed information is more desirable than a single 'optimal' estimate for the analysis and control of power systems. This method considers the measurement uncertainties of systems, which are described only through associated error boundaries, and does not assume that the measurement errors follow a probability density function. Therefore, the results are credible if the description of the measurement boundary is correct. Interval SE was originally developed for transmission systems [4].

*B. Related Literature*

The solutions for the interval SE model can be roughly classified into three types:

(1) Interval analysis (IA) can be applied to the model directly [7] and has been used in transmission networks [9]. However, the results generated by the IA are very conservative [10]. Modified Krawczyk-operator algorithm is proposed to solve interval linear state estimation based on PMU and SCADA hybrid measurements [11]. However, the deployment of PMUs in distribution networks is not affordable since their scales are much larger than transmission networks.

(2) The interval SE problem also can be formulated as optimization models [12]. Some researchers have shown that the uncertainty intervals of state variables and measurements could be estimated by programmatically maximizing or minimizing a variable component. However, the results are not guaranteed as the optimal problems are non-convex[6]. Two interval optimization models based on the unknown-but-bounded (UBB) theory and the solution bounds of state variables obtained using a two-stage linear programming (LP) approach were presented in Ref. [15] without considering shrinking the interval of pseudo measurement. For this nonlinear model, linearization can be performed in a certain system state, but this does not guarantee an optimal solution. A method for interval SE in the case of bad data was presented in Ref.[16], but the outcome of this technique relied on a scaling parameter.

(3) Interval constraint propagation (ICP) [17] is also used for interval SE. This is an efficient technique that has been applied for power system SE [5] [19]. Real-time measurements are insufficient, but large-scale pseudo-measurements can increase the flexibility of the ICP method. However, the ICP results are over-relaxed as the correlations of same interval variables in different terms of distribution network constraints are ignored.

*C. Contributions*

In this paper, we propose a linear programming contractor algorithm for DNSE in which the relative distance measure (RDM) interval operation [20] is applied. This method uses the computational framework of optimization model, but RDM interval calculations are used to estimate the optimal solution. The main contributions of this paper are as follows:

(1) A new scheme for interval SE is presented, in which the interval SE problem is formulated as an RDM based optimization model. Since RDM interval operations conform to laws in mathematical calculations (such as distributive law, cancellation law), the conservativeness of the estimation states can be significantly reduced compared to the conventional interval SE.

(2) The RDM based SE model is originally non-convex. To guarantee the credibility of the estimation results, the nonlinear measurement equations are transformed into an LP contractor. The LP contractor can iteratively narrow the upper and lower bounds of the system state variables. Additionally, the proposed method can shrink the bounds of pseudo measurements and the accuracy and credibility of estimation states is improved significantly.

(3) This RDM based SE model can address the three-phase unbalance problems in distribution networks, since the three-phase model is used.

The remainder of the paper is organized as follows. The main concepts of the RDM arithmetic are presented in Section II, and Section III describes the interval form of SE in distribution networks and its solution. A linear programming contractor algorithm using RDM interval arithmetic is proposed in Section IV, and Section V details the results of several numerical tests to justify the performance of the proposed algorithm. The paper is concluded in Section VI.

II. MAIN CONCEPTS OF RDM-ARITHMETIC

The conventional interval arithmetic does not consider the correlation between interval variables causes regular interval operations to not conform to some laws in mathematical calculations (such as distributive law, cancellation law)[17]. To solve this problem, the concept of multidimensional RDM arithmetic was developed by Piegat [22].

In RDM, the interval $X = [\underline{x}, \overline{x}]$ is described as:

$$X = \{x : x = \underline{x} + \alpha_x(\overline{x} - \underline{x}), \ \alpha_x \in [0,1]\} \qquad (1)$$

Here, the RDM variable $a_x$ makes it possible to obtain any value between the left and right boundaries of the interval X ($\underline{x}$ and $\bar{x}$, respectively). When $\alpha_x = 0$, the value from interval X equals $\underline{x}$ and $\alpha_x = 1$ gives $\bar{x}$. Figure 1 shows the interval $X = [\underline{x}, \bar{x}]$, and indicates the meaning of the RDM variable $a_x$ when $\underline{x} \leq \bar{x}$.

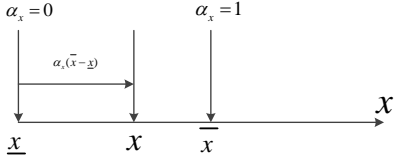

Figure 1 The interval $X = [\underline{x}, \bar{x}]$, and the meaning of the relative distance measurement (RDM) variable $\alpha_x \in [0,1]$.

### A. Operations in RDM interval arithmetic

The following operations are defined in RDM arithmetic: addition, subtraction, multiplication and division. Depending on the number of variables in a calculation, the obtained solutions are in multidimensional space, as opposed to one-dimensional space for conventional interval arithmetic.

Let $X$ and $Y$ represent two intervals:

$$X = [\underline{x}, \bar{x}] = \{x : x = \underline{x} + \alpha_x(\bar{x} - \underline{x}), \alpha_x \in [0,1]\} \quad (2)$$

$$Y = [\underline{y}, \bar{y}] = \{y : y = \underline{y} + \alpha_y(\bar{y} - \underline{y}), \alpha_y \in [0,1]\} \quad (3)$$

*1) Addition in RDM*

$$X + Y = \{x + y : x + y = \underline{x} + \alpha_x(\bar{x} - \underline{x}) + \underline{y} + \alpha_y(\bar{y} - \underline{y})\}$$
$$\alpha_x, \alpha_y \in [0,1]\} \quad (4)$$

*2) Subtraction in RDM*

$$X - Y = \{x - y : x + y = \underline{x} + \alpha_x(\bar{x} - \underline{x}) - \bar{y} - \alpha_y(\bar{y} - \underline{y})\}$$
$$\alpha_x, \alpha_y \in [0,1]\} \quad (5)$$

*3) Multiplication in RDM*

$$X \times Y = \{xy : xy = [\underline{x} + \alpha_x(\bar{x} - \underline{x})] \times [\underline{y} + \alpha_y(\bar{y} - \underline{y})]\}$$
$$\alpha_x, \alpha_y \in [0,1]\} \quad (6)$$

*4) Division in RDM*

$$X / Y = \{x - y : x + y = [\underline{x} + \alpha_x(\bar{x} - \underline{x})] / [\underline{y} + \alpha_y(\bar{y} - \underline{y})]\}$$
$$\alpha_x, \alpha_y \in [0,1]\} \quad (7)$$

For intervals $X = [x^-, x^+]$ and $Y = [y^-, y^+]$, the base operations $* \in \{+, -, \times, /\}$ span is an interval defined as (8); operation / is defined only if $0 \notin Y$.

$$s(X * Y) = [\min(X * Y), \max(X * Y)] \quad (8)$$

Operations in RDM interval arithmetic conform to all properties of general algebra: commutativity, associativity, inverse elements, distributive law, the cancellation law of multiplication, and so on. But the conventional interval arithmetic cannot fully meet these properties. Further details are provided in Appendix A of [25] as a supplemental file.

### B. Correlation analysis of variables in RDM arithmetic

Taking $f = x - x^2$ with an interval $[x] = [1,2]$ as an example, we can test the above assertions. Set $b = x^2; c = 1 - x; d = 1 + x$. Then we have the function calculation table:

| Function form | Conventional interval arithmetic | RDM interval arithmetic |
|---|---|---|
| $f_1 = x - b$ | $f_1 = [-3, 1]$ | $f_1 = [-2, 0]$ |
| $f_2 = x \times c$ | $f_2 = [-2, 0]$ | $f_2 = [-2, 0]$ |
| $f_3 = -1 + x + c \times d$ | $f_3 = [-3, 1]$ | $f_3 = [-2, 0]$ |

We know that $f_1 = f_2 = f_3$ in mathematics, but the conventional interval arithmetic obtains different results. This occurs because the conventional interval arithmetic consider a, b, and c are independent variables without considering their correlations. For conventional interval arithmetic, the results depend on the specific function form and may be very conservative. In contrast, the RDM interval arithmetic fixes this defect.

**Remark**: RDM interval arithmetic can eliminate the conservativeness of the correlation problem when some interval variables appear several times. Furthermore, the RDM interval algorithm does not depend on the function form and conforms to all of the properties of mathematical calculation theory.

## III. INTERVAL FORM OF STATE ESTIMATION IN DISTRIBUTION NETWORKS

### A. Deficiencies in traditional state estimation

The relationship between the state variables and measurements is determined by the measurement equations, as:

$$Z = h(x) + v \quad (9)$$

where $Z \in R^m$ are the measurement vectors, $x \in R^n$ is the state vector, $e \in R^m$ is the measurement error vector, and $h : R^n \to R^n$ are the non-linear measurement functions. There are equality constraints for zero injection nodes:

$$c(x) = 0 \quad (10)$$

Thus, the SE model can be formulated as:

$$\begin{aligned} \min \quad & f(Z - h(x)) \\ s.t \quad & c(x) = 0 \end{aligned} \quad (11)$$

The result of (11) depends on the estimation criteria used (e.g. weighted least squares estimator, quadratic-constant estimator and so on), but in these existing methods, measurements with uncertain errors are described as random variables with known statistical properties, and the estimations are handled using probability theory. In practice, these statistical properties are difficult to characterize. Inexact matching of the assumed statistical hypotheses may lead to inaccurate estimates. Therefore, the interval state estimation becomes an effective alternative method.

### B. Interval form of state estimation

The interval state estimation consider that the measurement error ($v \in R^m$) is bounded within a known range and can therefore be expressed as:

$$E = \{v \in R^m \mid \underline{v} \leq v \leq \bar{v}\} \quad (12)$$

Where $\underline{v}$ and $\bar{v}$ are the prior bounds of the error set, and $\underline{v} \leq 0 \leq \bar{v}$.

In the bounded-error context, each measurement equation can be represented by upper and lower limits as:

$$Z - \bar{v} \le h(x) \le Z - \underline{v} \quad (13)$$

In this context, the measurement equation can be expressed as:

$$Z - h(x) = 0 \quad (14)$$

Where $Z \in [Z - \bar{v}, Z - \underline{v}]$ is an estimation of the quantities in the measurement vector $Z$.

According to the interval SE approach, the estimated states are denoted as intervals $[x]$, which is the solution of the following constrained satisfaction problem:

$$S = \{[x] \in [x^0], Z \in [Z - \bar{v}, Z - \underline{v}] \mid Z - h([x]) = 0, c([x]) = 0\} \quad (15)$$

Where $[x^0]$ is the initial interval value of the state variables and set $S$ provides an accurate description of the state uncertainty. The interval SE guarantees suitable estimates of the of characteristics of $S$.

It is clear that if the true value $x_t$ of the state variable satisfies $c(x_t) = 0$ and the error set is accurately described, then $x_t \in [x]$ (i.e., if $Z - h(x_t) \in E$ then $x_t \in [x]$). This shows the credibility of the set $[x]$.

## IV. RDM FORM OF INTERVAL STATE ESTIMATION FORMULATION AND ITS SOLUTION

In this paper, the real and imaginary parts of voltage and the branch current are chosen as the state variables as listed in (16). Therefore, measurement functions of ampere and voltage measurements are significantly simplified, and no measurements transformation is needed in this method.

$$x = (e_i, f_i, I_{ij}^{re}, I_{ij}^{im}) \quad (16)$$

SE measurement variables in the distribution network include the square of branch current $I_{ij}^m$, the square of voltage $V_i^m$, active and reactive injection power $P_i^m, Q_i^m$, and branch power $P_{ij}^{(m)}, Q_{ij}^{(m)}$

$$Z = (V_i^{(m)}, P_i^{(m)}, Q_i^{(m)}, I_{ij}^{(m)}, P_{ij}^{(m)}, Q_{ij}^{(m)}) \quad (17)$$

### A. Measurement and Constraint Equations for RDM Intervals State Estimation of Distribution Networks

In interval SE, all the variables and measurements can be expressed as intervals：

$$[x] = [\underline{x}, \bar{x}], [Z] = [\underline{Z}, \bar{Z}] \quad (18)$$

The measurement and constraint equations of the DNSE are described as followings (all variables and measurements are intervals).

*1) The squared branch current amplitude measurement equation*

$$L_{ij(s)}^{(m)} = (I_{ij(s)}^{(m)})^2 = (I_{ij(s)}^{re})^2 + (I_{ij(s)}^{im})^2 + (v_{L_{ij(s)}}) \quad (19)$$

(19) can be reformulated into interval form:

$$[L_{ij(s)}^{(m)}] = [I_{ij(s)}^{re}]^2 + [I_{ij(s)}^{im}]^2 \quad (20)$$

Their RDM interval form are:

$$(L_{ij(s)}^{(m)})_{RDM} = \underline{L_{ij(s)}^{(m)}} + \alpha_{ij(s)}^{L(m)}(\overline{L_{ij(s)}^{(m)}} - \underline{L_{ij(s)}^{(m)}})$$

$$(I_{ij(s)}^{re})_{RDM} = \underline{I_{ij(s)}^{re}} + \alpha_{ij(s)}^{Ire(s)}(\overline{I_{ij(s)}^{re}} - \underline{I_{ij(s)}^{re}}) \quad (21)$$

$$(I_{ij(s)}^{im})_{RDM} = \underline{I_{ij(s)}^{im}} + \alpha_{ij(s)}^{lim}(\overline{I_{ij(s)}^{im}} - \underline{I_{ij(s)}^{im}})$$

Equation (20) can be rearranged as RDM interval arithmetic form：

$$\underline{L_{ij(s)}^{(m)}} - (\underline{I_{ij(s)}^{re}})^2 - (\underline{I_{ij(s)}^{im}})^2$$
$$= -\alpha_{ij(s)}^{L(m)}(\overline{L_{ij(s)}^{(m)}} - \underline{L_{ij(s)}^{(m)}}) + 2\alpha_{ij(s)}^{Ire} \underline{I_{ij(s)}^{re}}(\overline{I_{ij(s)}^{re}} - \underline{I_{ij(s)}^{re}})$$
$$+ (\alpha_{ij(s)}^{Ire})^2(\overline{I_{ij(s)}^{re}} - \underline{I_{ij(s)}^{re}})^2 + 2\alpha_{ij(s)}^{lim} \underline{I_{ij(s)}^{im}}(\overline{I_{ij(s)}^{im}} - \underline{I_{ij(s)}^{im}}) \quad (22)$$
$$+ (\alpha_{ij(s)}^{lim})^2(\overline{I_{ij(s)}^{im}} - \underline{I_{ij(s)}^{im}})^2$$

Other measurement equations can also be reformulated into RDM interval arithmetic form.

*2) The power balance constraints on buses*

The three-phase the branch current constraints are:

$$I_{ij(s)}^{re} = I_{j(s)}^{re} + \sum_{k \in j} I_{jk(s)}^{re}$$
$$I_{ij(s)}^{im} = I_{j(s)}^{im} + \sum_{k \in j} I_{jk(s)}^{im} \quad (23)$$

Then,

$$I_{j(s)}^{re} = \frac{P_{j(s)}^{(m)} e_{j(s)} + Q_{j(s)}^{(m)} f_{j(s)}}{e_{j(s)}^2 + f_{j(s)}^2}$$
$$I_{j(s)}^{im} = \frac{P_{j(s)}^{(m)} f_{j(s)} - Q_{j(s)}^m e_{j(s)}}{e_{j(s)}^2 + f_{j(s)}^2} \quad (24)$$

From (23) and (24), the power balance constraints on buses are formulated as:

$$0 = \frac{P_{j(s)}^{(m)} e_{j(s)} + Q_{j(s)}^{(m)} f_{j(s)}}{V_{j(s)}} - I_{ij(s)}^{re} + \sum_{k \in j} I_{jk(s)}^{re}$$
$$0 = \frac{P_{j(s)}^{(m)} f_{j(s)} - Q_{j(s)}^m e_{j(s)}}{V_{j(s)}} - I_{ij(s)}^{im} + \sum_{k \in j} I_{jk(s)}^{im} \quad (25)$$

Where ： $V_{j(s)} = e_{j(s)}^2 + f_{j(s)}^2$.

(25) can be reformulated into interval form:.

$$0 = \frac{[P_{j(s)}^{(m)}][e_{j(s)}] + [Q_{j(s)}^{(m)}][f_{j(s)}]}{[V_{j(s)}]} - [I_{ij(s)}^{re}] + \sum_{k \in j}[I_{jk(s)}^{re}]$$
$$0 = \frac{[P_{j(s)}^{(m)}][f_{j(s)}] - [Q_{j(s)}^m][e_{j(s)}]}{[V_{j(s)}]} - [I_{ij(s)}^{im}] + \sum_{k \in j}[I_{jk(s)}^{im}] \quad (26)$$

And RDM arithmetic form：

$$\underline{I_{ij(s)}^{re}} - \sum_{k \in j} \underline{I_{jk(s)}^{re}} = \frac{(P_{j(s)}^{(m)})_{RDM}(e_{j(s)})_{RDM} + (Q_{j(s)}^{(m)})_{RDM}(f_{j(s)})_{RDM}}{\underline{V_{j(s)}} + \alpha_{j(s)}^V(\overline{V_{j(s)}} - \underline{V_{j(s)}})}$$
$$- \alpha_{ij(s)}^{Ire}(\overline{I_{ij(s)}^{re}} - \underline{I_{ij(s)}^{re}}) + \sum_{k \in j} \alpha_{jk(s)}^{Ire}(\overline{I_{jk(s)}^{re}} - \underline{I_{jk(s)}^{re}})$$

$$\underline{I_{ij(s)}^{im}} - \sum_{k \in j} \underline{I_{jk(s)}^{im}} = \frac{(P_{j(s)}^{(m)})_{RDM}(f_{j(s)})_{RDM} - (Q_{j(s)}^{(m)})_{RDM}(f_{j(s)})_{RDM}}{\underline{V_{j(s)}} + \alpha_{j(s)}^V(\overline{V_{j(s)}} - \underline{V_{j(s)}})} \quad (27)$$
$$- \alpha_{ij(s)}^{lim}(\overline{I_{ij(s)}^{im}} - \underline{I_{ij(s)}^{im}}) + \sum_{k \in j} \alpha_{jk(s)}^{lim}(\overline{I_{jk(s)}^{im}} - \underline{I_{jk(s)}^{im}})$$

Where:

$$(P_{j(s)}^{(m)})_{RDM} = \underline{P_{j(s)}^{(m)}} + \alpha_{j(s)}^{P(m)}(\overline{P_{j(s)}^{(m)}} - \underline{P_{j(s)}^{(m)}})$$

$$(Q_{j(s)}^{(m)})_{RDM} = \underline{Q_{j(s)}^{(m)}} + \alpha_{j(s)}^{Q(m)}(\overline{Q_{j(s)}^{(m)}} - \underline{Q_{j(s)}^{(m)}})$$

$$(e_{j(s)})_{RDM} = \underline{e_{j(s)}} + \alpha_{j(s)}^{e}(\overline{e_{j(s)}} - \underline{e_{j(s)}})$$

$$(f_{j(s)})_{RDM} = \underline{f_{j(s)}} + \alpha_{j(s)}^{e}(\overline{f_{j(s)}} - \underline{f_{j(s)}})$$
(28)

3) *The measurement equation for the squared voltage is*

$$V_{j(s)}^{(m)} = (U_{j(s)}^{m})^2 = e_{j(s)}^2 + f_{j(s)}^2 + v_{(U_{j(s)}^m)^2}$$ (29)

It can be reformulated into interval form:

$$[V_{j(s)}^{(m)}] = [e_{j(s)}^2] + [f_{j(s)}^2]$$ (30)

And RDM interval arithmetic form:

$$\underline{V_{j(s)}^{(m)}} - (\underline{e_{j(s)}})^2 - (\underline{f_{j(s)}})^2$$
$$= -\alpha_{j(s)}^{V(m)}(\overline{V_{j(s)}^{(m)}} - \underline{V_{j(s)}^{(m)}}) + 2\alpha_{j(s)}^{e}\underline{e_{j(s)}}(\overline{e_{j(s)}} - \underline{e_{j(s)}})$$
$$+ 2\alpha_{j(s)}^{f}\underline{f_{j(s)}}(\overline{f_{j(s)}} - \underline{f_{j(s)}}) + (\alpha_{j(s)}^{e})^2(\overline{e_{j(s)}} - \underline{e_{j(s)}})$$
$$+ (\alpha_{j(s)}^{f})^2(\overline{f_{j(s)}} - \underline{f_{j(s)}})$$
(31)

4) *The voltage constraint of node is given by*

$$e_{i(s)} - e_{j(s)} - r_{ij(s)(p)}I_{ij(p)}^{re} + x_{ij(s)(p)}I_{ij(p)}^{im} = 0$$
$$f_{i(s)} - f_{j(s)} - x_{ij(s)(p)}I_{ij(p)}^{re} - r_{ij(s)(p)}I_{ij(p)}^{im} = 0$$
(32)

It can be further reformulated into interval form:

$$[e_{i(s)}] - [e_{j(s)}] - r_{ij(s)(p)}[I_{ij(p)}^{re}] + x_{ij(s)(p)}[I_{ij(p)}^{im}] = 0$$
$$[f_{i(s)}] - [f_{j(s)}] - x_{ij(s)(p)}[I_{ij(p)}^{re}] - r_{ij(s)(p)}[I_{ij(p)}^{im}] = 0$$
(33)

And RDM interval arithmetic form:

$$\underline{e_{j(s)}} - \underline{e_{i(s)}} + r_{ij(s)(p)}\underline{I_{ij(p)}^{re}} - x_{ij(s)(p)}\underline{I_{ij(p)}^{im}}$$
$$= \alpha_{i(s)}^{e}(\overline{e_{i(s)}} - \underline{e_{i(s)}}) - \alpha_{j(s)}^{e}(\overline{e_{j(s)}} - \underline{e_{j(s)}})$$
$$- r_{ij(s)(p)}\alpha_{ij(p)}^{Ire}(\overline{I_{ij(p)}^{re}} - \underline{I_{ij(p)}^{re}}) + x_{ij(s)(p)}\alpha_{ij(p)}^{Iim}(\overline{I_{ij(p)}^{im}} - \underline{I_{ij(p)}^{im}})$$
$$\underline{f_{j(s)}} - \underline{f_{i(s)}} + x_{ij(s)(p)}\underline{I_{ij(p)}^{re}} + r_{ij(s)(p)}\underline{I_{ij(p)}^{im}}$$
$$= \alpha_{i(s)}^{f}(\overline{f_{i(s)}} - \underline{f_{i(s)}}) - \alpha_{j(s)}^{f}(\overline{f_{j(s)}} - \underline{f_{j(s)}})$$
$$- x_{ij(s)(p)}\alpha_{ij(p)}^{Ire}(\overline{I_{ij(p)}^{re}} - \underline{I_{ij(p)}^{re}}) - r_{ij(s)(p)}\alpha_{ij(p)}^{Iim}(\overline{I_{ij(p)}^{im}} - \underline{I_{ij(p)}^{im}})$$
(34)

Here, $p \in s$ denote the phase indices.

The variables in the RDM measurement equations can be classified into two types. The first type is the RDM variables of the states.

$$\alpha_{(s)}^{x} = (\alpha_{j(s)}^{e}, \alpha_{j(s)}^{f}, \alpha_{ij(s)}^{Ire}, \alpha_{ij(s)}^{Iim})$$ (35)

The second type is the RDM variables of the measurements:

$$\alpha_{(s)}^{Z(m)} = (\alpha_{j(s)}^{P(m)}, \alpha_{j(s)}^{Q(m)}, \alpha_{j(s)}^{L(m)}, \alpha_{j(s)}^{V(m)})$$ (36)

Obviously, all these two types of variables can be optimized. Since the uncertainties of pseudo measurements are initially overestimated, i.e. the initial range of their interval are conservative. (36) indicates that the measurement range can be described as the RDM variables, the final range of measurements can be narrowed according to the constraints of this interval SE model.

### B. Linear programming contractor

As (22), (27) and (28), (31) are nonlinear, we present a linearization strategy (referred to as an LP contractor) to transform the model into convex form. The specific process is described in Appendix B of [25].

The nonlinear constraints (22), (27) and (28), (31) can be rewritten in a compact form:

$$g(x) = 0$$ (37)

As $g$ is differentiable, according to the mean-value theorem (22), (27) and (28), (31) can be transformed into linear inequality constraints as a contractor.

$$\begin{pmatrix} -A \\ A \end{pmatrix} \alpha_x \leq \begin{pmatrix} -\underline{B} \\ \overline{B} \end{pmatrix}$$ (38)

Where:

$$A = J_g(\underline{x_0})(\overline{x_0} - \underline{x_0})$$
$$B = [\underline{B}, \overline{B}] = (-g(\underline{x_0}) - (J_g([x]) - J_g(\underline{x_0}))(\overline{x_0} - \underline{x_0})\alpha_x)$$
$$J_g(x) = \frac{\partial g(x)}{\partial x}; \underline{x_0} = \inf([x]); \overline{x_0} = \sup([x])$$
$$[x] = ([P_{j(s)}^{(m)}], [Q_{j(s)}^{(m)}], [L_{ij(s)}^{(m)}], [V_{j(s)}^{(m)}], [e_{j(s)}], [f_{j(s)}], [I_{ij(s)}^{re}], [I_{ij(s)}^{im}])$$
$$\alpha_x = [\alpha_{j(s)}^{P(m)} \ \alpha_{j(s)}^{Qm} \ \alpha_{ij(s)}^{L(m)} \ \alpha_{j(s)}^{V(m)} \ \alpha_{j(s)}^{e} \ \alpha_{j(s)}^{f} \ \alpha_{ij(s)}^{Ire} \ \alpha_{ij(s)}^{Iim}]^T$$

The boundaries of $[x]$ narrow during the iteration process, so $B$ are also reduced. Therefore, Equations (38) can iteratively reduce $[x]$ as an LP contractor and the whole algorithm procedure is described in C section.

### C. Solution for the Interval State Estimation

The solution procedure for the proposed SE model as follows:

**Step 1:** Set $t=0$, initialize measurements $[Z^t] = [\underline{Z^t}, \overline{Z^t}]$ and state variables $[X^t] = [\underline{X^t}, \overline{X^t}]$. Convert the original interval equations into RDM form.

**Step 2:** Turn the problems into LP-RDM issues by reference to section A by solving

$$\min \text{ or } \max \ \alpha(\alpha_{j(s)}^{P(m)}, \alpha_{j(s)}^{Qm}, \alpha_{ij(s)}^{L(m)}, \alpha_{j(s)}^{V(m)}, \alpha_{j(s)}^{e}, \alpha_{j(s)}^{f}, \alpha_{ij(s)}^{Ire}, \alpha_{ij(s)}^{Iim})$$
$$s.t \ (22),(27),(28),(31),(34),(38)$$
$$0 \leq \alpha_{j(s)}^{P(m)}, \alpha_{j(s)}^{Qm}, \alpha_{ij(s)}^{L(m)}, \alpha_{j(s)}^{V(m)}, \alpha_{j(s)}^{e}, \alpha_{j(s)}^{f}, \alpha_{ij(s)}^{Ire}, \alpha_{ij(s)}^{Iim} \leq 1$$
(39)

The basic idea of the model of (39) is: Since the RDM variable monotonically increases within a range (explained in section II), as long as the min and max of the RDM variable are solved, the range of the original interval can be calculated. The significance of this model is: when specifying any point in each measurement range, the corresponding state variable point can be found, ensuring that all possible states of the model can be found, avoiding the impact of uncertainty.

**Step 3**: Convert the RDM variables into their original interval forms, update the measurements $[Z^{t+1}]$ and state variables $[X^{t+1}]$.

It should be noted that each time the RDM variable is optimized, the upper and lower bounds of the measurement interval and the upper and lower bounds of the variable are changed.

**Step 4**: If the widths of $[Z^{t+1}]$ and $[X^{t+1}]$ are not decreased further, go to **Step 5**; otherwise $t=t+1$, go to **Step 2.**

**Step 5**: End

## V. SIMULATION RESULT

The proposed algorithm was implemented in the MATLAB environment and the INTLAB package [23] was employed for IA computation. For comparison, we implemented two alternative techniques: the conventional ICP method[5], Modified Krawczyk-operator [11] and an optimization-based interval method (OPT). The detailed model of OPT method can refer to Appendix D of [25] as a supplemental file.

The tests were divided into three parts:
- Conservative test: The SE interval widths were considered, and the width of the state interval was defined as:

$$wid_{avr}([x]) = \sum_{i=1}^{n} w([x_i]) / n \quad (40)$$

For the measurement interval $[z]$, the initial interval $[z^0]$ was known, so the degree of conservativeness could be evaluated by comparing the measurement interval shrinkage ratios:

$$ratio([z]) = \sum_{j=1}^{m} wid([z]) / wid([z^0]) \quad (41)$$

- Credibility test: a credibility test was used to determine how frequently the true value was within the resulting interval. The calculation index $C$ is defined as:

$$C = \frac{1}{N}\sum_{i=1}^{N} c_i \quad (42)$$

Where $N$ is the total number of test samples,

$$c_i = \begin{cases} 1 & if \ C_{1i} \geq 0.95 \ and \ C_{2i} \geq 0.95 \\ 0 & otherwise \end{cases}$$

And

$$C_1 = \frac{1}{n}\sum_{i=1}^{n} c_{1i}, \quad C_2 = \frac{1}{m}\sum_{i=1}^{m} c_{2i} \quad (43)$$

Where $C_1$ is the confidence in the state variable, $C_2$ is the confidence in the measurement variable, $n$ is the number of state variables, and $m$ is the number of measurements. $c_{1i} = 1$ if $x_{true} \in [\underline{x_i}, \overline{x_i}]$; otherwise, $c_{1i} = 0$; $x_{true}$ is the true value of the state variable. $c_{2i} = 1$ if $z_{true} \in [\underline{z_i}, \overline{z_i}]$; otherwise, $c_{2i} = 0$; $z_{true}$ is the true value of the measurement variable.

- Computational performance: Additional tests have been performed to assess the computational performance of the proposed method.

All tests were carried out on IEEE 33-bus and 123-bus networks, and the systems were configured as described in [24]. In total, 100 measurements were simulated by adding uniformly distributed noise to a real set of measured data. The uncertainty variables considered here include voltage, current, and power. The measurement details are listed in [24]. In practice, the uncertainty of measurement are various. The uncertainty variables considered in this paper include voltage, current and power data. The standard deviations for each type are listed as follows:

✓ Initial voltage width $[0.9, 1.1]$

✓ Due to the characteristics of the distribution network, the phase angle can be set to $[-5^o, 5^o]$

✓ The branch current variable can be obtained according to the method in [11].

✓ For the measurement data, the standard deviation of current measurement is 1%. The standard deviations of active and reactive power measurement are 1%. So, the initial measurement interval was $[0.99z, 1.01z]$. z is the measurement value.

As pseudo- measurement errors are large, the upper and lower ranges of pseudo-measurement fluctuations were set to roughly ±10% of their rated values [10].

### A. IEEE 33-bus distribution network

A credible system is one in which the result contains all solutions that satisfy the constraints, and reflect the true system state. The true state of a standard test system is known, so it is possible to directly test whether the true state was included in the result interval. Here, 100 samples on the IEEE 33-bus network were simulated by adding a uniformly distributed error to the true measurement values. Table I shows that the proposed method and original ICP method obtained guaranteed results in all cases (100%), but the OPT method received a credibility score of 67%.

Table I show the widths determined by the three methods and the node voltage amplitude interval, which is determined using equation (40-41) and (42-43). It was found that the smallest width was estimated using the OPT method, but the credibility of the method is not guaranteed. The proposed method is clearly superior to the ICP method and modified Krawczyk operator (MKO) [11] because it does not suffer from the correlation problem. Therefore, the conservativeness of the results obtained by the proposed method is significantly reduced compared to those of the ICP method and MKO. Table I also lists the calculation time of the three methods for the 33-node system. The proposed method was 80 times faster than the OPT method and also a little faster than the ICP method.

TABLE I
RESULTS OBTAINED WITH THE IEEE 33-BUS NETWORK

| Method | Credibility | Voltage Average width | Ratio | Calculating time/s |
|---|---|---|---|---|
| OPT | 67 | 0.00064 | 0.7513 | 5918.39 |
| ICP | 100 | 0.0034 | 1 | 80.09 |
| MKO | 100 | 0.0042 | 1 | 36.7250 |
| Proposed | 100 | 0.0009 | 0.9236 | 71.56 |

Table I also presents the shrinkage ratios for the measurement. It shows that the conservativeness of the proposed method is reduced compared to the ICP method. Specifically, the ratio obtained by the ICP method and MKO is close to 1, while for the proposed method it is 0.9236. It can be seen that proposed method can effectively mitigate the conservativeness associated with the ICP method and MKO method. It should be noted that the OPT method relies heavily on the initial width and value selections, and it will fail to converge if the initial value is improperly selected. Fig.2 shows magnitude results of the phase in the 33-bus system.

As can be seen from fig.2, the results of this paper show the best results at each phase voltage, and the results of the proposed method contain truth values, indicating that the results are credible.

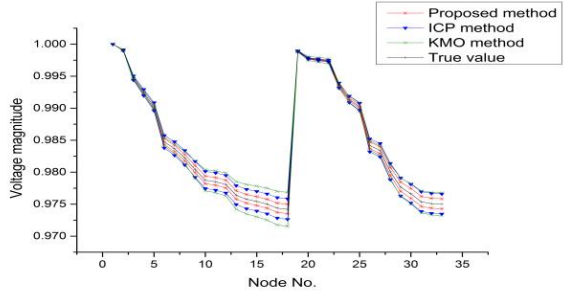

(a) Phase A

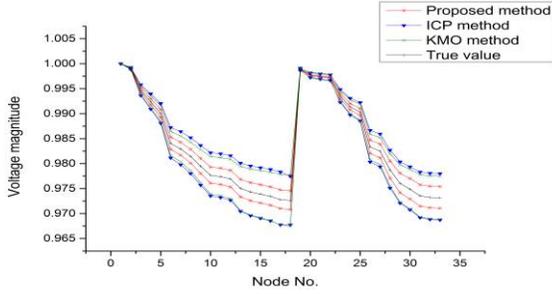

(b) Phase B

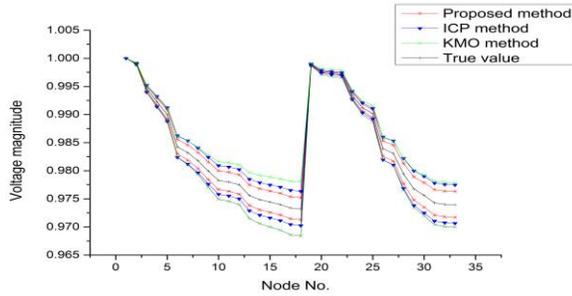

(c) Phase C

Fig.2 Voltage magnitude results in the 33-bus system

Table 2 lists the width of the system current branch variable and the overall width of the system. It can be seen from Table 2 that the results of the OPT method are the best, but the OPT method is not credible according to the above analysis. Therefore, the proposed method is the best one among credible solutions.

TABLE II
RESULTS OBTAINED WITH THE IEEE 33-BUS NETWORK

| Method | Real part width of branch current | Imaginary part width of current branch | overall width of the system |
|---|---|---|---|
| OPT | 0.0012 | 0.0010 | 0.0009 |
| ICP | 0.0046 | 0.0042 | 0.0037 |
| MKO | 0.0091 | 0.0079 | 0.0075 |
| Proposed | 0.0037 | 0.0033 | 0.0024 |

Fig.4 shows the voltage average width and the measured width shrink decrease monotonously during the iteration procedure of the proposed method. This means the proposed method can narrow the voltage width and the measurement width as the iteration proceeding, which makes the inequality constraints tighter, forming a contraction process. After the iteration number reaches 5, the voltage width and the measurement width are basically unchanged, which indicates that the contraction process converges.

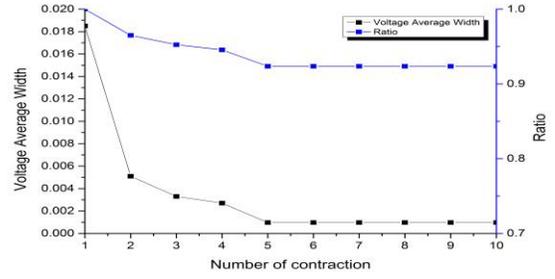

Fig.4 The number of contractions in the calculation process

*B. IEEE 123-bus distribution network*

Similar to the IEEE 33-bus network, the credibility of the OPT method was only 62%, while the other two methods were 100% credible. Table III lists the average voltage width and measurement shrinking ratio values. The OPT method produced the best results, but it has been shown to have poor credibility. The proposed method obtained significantly better results than the ICP method and MKO. Table III also shows that the proposed method is more efficient than the OPT method and ICP method.

TABLE III
RESULTS OBTAINED WITH THE IEEE 123-BUS NETWORK

| Method | Credibility | Average voltage width | Ratio | Calculating time/s |
|---|---|---|---|---|
| OPT | 62 | 0.0017 | 0.4732 | 29767.91 |
| ICP | 100 | 0.0088 | 0.9074 | 364.34 |
| MKO | 100 | 0.0278 | 1 | 123.54 |
| Proposed | 100 | 0.0033 | 0.8267 | 259.52 |

Table IV lists the estimation results of each phase voltage. The proposed method shows the best performance in any aspect from the table (Re: Real part, Im: Imaginary part). The MKO method estimated the worst results, this is because it cannot shrink measurement width. Although the ICP method can shrink measurement with to some extent, it ignores the correlation between variables, which leads to very conservative results. The proposed method can not only avoid the correlation problem of the ICP method, but also can shrinking the measurement width, so it has the potential to be applied in the real distribution networks.

TABLE IV
AVERAGE VOLTAGE WIDTH OF RESULTS OBTAINED WITH THE IEEE 123-BUS NETWORK

| Method | A phase | | B phase | | C phase | |
|---|---|---|---|---|---|---|
| | Re | Im | Re | Im | Re | Im |
| OPT | ----- | ----- | ------ | ----- | ----- | ----- |
| ICP | 0.008 | 0.008 | 0.010 | 0.011 | 0.009 | 0.007 |
| MKO | 0.024 | 0.016 | 0.031 | 0.033 | 0.031 | 0.032 |
| Proposed | 0.003 | 0.003 | 0.003 | 0.004 | 0.004 | 0.003 |

Table V lists the details of shrinking the pseudo-measurement width in various methods. It shows that the initial width of the measurement can be set wide for safely in our method, because it can iteratively shrink the pseudo-measurement width and improve the estimation accuracy

TABLE V
PSEUDO-MEASUREMENT OF SHRINKAGE RATIO

| Method | A phase | | B phase | | C phase | |
|---|---|---|---|---|---|---|
| | Active power | Reactive power | Active power | Reactive power | Active power | Reactive power |
| OPT | ----- | ----- | ------ | ----- | ----- | ----- |
| ICP | 0.833 | 0.967 | 1 | 1 | 0.856 | 1 |
| MKO | 1 | 1 | 1 | 1 | 1 | 1 |
| Proposed | 0.767 | 0.967 | 0.806 | 1 | 0.769 | 1 |

## VI. CONCLUSION

Because the conventional interval SE methods overlook the correlations between different terms with same interval variables in constraints, the estimation results are overly conservative. We proposed a novel interval SE model by introducing RDM arithmetic, and the conservativeness of the interval SE method was reduced through inclusion of the RDM interval algorithm. Then, we proposed an efficient LP contractor solve the proposed SE model, it can iteratively shrink the state and measurement intervals. Furthermore, an OPT method was developed for comparison. Numerical tests showed that the OPT method produced the narrowest interval state width, but its credibility was unacceptably low. The proposed method guaranteed the credibility of the estimation results, and it also significantly mitigated the conservativeness compared to the conventional ICP and MKO methods.

■